\documentclass[a4paper,11pt]{article}
\usepackage{pos}
\usepackage{amsmath}
\usepackage{array}
\usepackage{booktabs}

\title{Primordial Black Holes and Gravitational Waves in Extensions of the Standard Model}
\author[a]{Indra Kumar Banerjee,}
\author[b]{Ujjal Kumar Dey,}
\author*[c]{Shaabn Khalil}

\affiliation[a]{Department of Physical Sciences, Indian Institute of Science Education and Research Berhampur,\\Ganjam 760003, Odisha, India}
\affiliation[b]{Center for Fundamental Physics, \\
Zewail City of Science and Technology, 6th of October City, Giza
12578, Egypt}

\emailAdd{indrab@iiserbpr.ac.in}
\emailAdd{ujjal@iiserbpr.ac.in}
\emailAdd{skhalil@zewailcity.edu.eg}

\abstract{We investigate the phenomenology of a Standard Model extension incorporating an inert scalar doublet and a gauged \( U(1)_{B-L} \) symmetry. Our analysis reveals regions of the parameter space that support strong first-order phase transitions, including cases featuring two successive transitions. Each transition can generate a stochastic gravitational wave background within the sensitivity reach of upcoming experiments. Remarkably, the high-scale transition may also produce primordial black holes with appreciable abundance.}

\FullConference{Proceedings of the Corfu Summer Institute 2024 "School and Workshops on Elementary Particle Physics and Gravity" (CORFU2024)\
12 - 26 May, and 25 August - 27 September, 2024\\
Corfu, Greece\\}

\begin{document}
\maketitle
\section{Introduction}

Primordial Black Holes (PBHs) and Gravitational Waves (GWs) stand as profound cosmic messengers, offering unique insights into the earliest moments of our universe. Unlike astrophysical black holes, PBHs may have formed shortly after the Big Bang, carrying imprints of high-energy processes and physics beyond the Standard Model (SM). Their existence and properties serve as invaluable probes into the dynamics of the early universe, its phase transitions, and the nature of dark matter.

PBHs can originate from the collapse of large density perturbations, potentially seeded during cosmic inflation. Among the various formation mechanisms, strong first-order phase transitions are particularly intriguing. These transitions are characterized by the nucleation and collision of expanding true-vacuum bubbles, which not only can induce sufficient over-densities to produce PBHs but also generate a stochastic background of gravitational waves. Such a dual signature makes them an exciting avenue for both theoretical and observational exploration.

The concept of PBHs was first introduced in the 1960s by Zel'dovich and Novikov~\cite{Zeldovich:1967lct}, and later developed further by Hawking and Carr in the 1970s~\cite{Carr:1974nx}. They proposed that PBHs could form from high-density regions in the early universe, in contrast to stellar-origin black holes formed from the collapse of massive stars. Despite decades of interest, direct detection of PBHs remains elusive. However, a range of indirect observational strategies has been proposed, including gravitational microlensing surveys~\cite{Niikura:2017zjd, Griest:2013esa, Niikura:2019kqi, Macho:2000nvd, EROS-2:2006ryy, Zumalacarregui:2017qqd} and their potential imprints on the cosmic microwave background~\cite{Poulin:2017bwe}.

The landmark detection of gravitational waves by LIGO and Virgo~\cite{LIGOScientific:2016aoc} has reignited interest in PBHs, especially in light of the possibility that some observed mergers could involve primordial rather than astrophysical black holes. Additionally, recent results from Pulsar Timing Arrays (PTAs) have provided compelling evidence for a stochastic gravitational wave background at nano-Hertz frequencies, potentially originating from early-universe sources such as phase transitions or cosmic strings.

In this work, we focus on a well-motivated extension of the Standard Model featuring an inert scalar doublet and a gauged \( U(1)_{B-L} \) symmetry. We demonstrate that this setup can support strong first-order phase transitions, which not only yield detectable gravitational wave signatures but also facilitate the formation of PBHs in the early universe. Our results highlight the rich interplay between particle physics and cosmology, offering a viable framework where PBHs and GWs emerge as correlated phenomena tied to high-scale physics beyond the SM. This presentation is based on the recent article by Banerjee et al.~\cite{Banerjee:2024fam}, where these ideas are developed in greater detail.

%%%%%%%%%%%%%%%%%%%%%%%%%%%%%%%%%%
\section{Extension of the SM}

The extension of the SM proposed here introduces an additional $U(1)_{B-L}$ gauge symmetry, where $B-L$ represents the difference between the baryon number $B$ and lepton number $L$. This symmetry is motivated by several theoretical considerations, such as providing a natural framework for understanding neutrino masses through the seesaw mechanism, as well as offering a path to understanding the baryon asymmetry of the universe. The extended gauge group of the model is:
\begin{equation}
G = SU(3)_C \times SU(2)_L \times U(1)_Y \times U(1)_{B-L}
\end{equation}

In this extension, the new $U(1)_{B-L}$ symmetry is an exact gauge symmetry of the model, and the fields of the theory transform accordingly. The relevant field content of the model is summarized in Table \ref{tab:BSM_fields}. This table outlines the quantum numbers of the fields under the extended gauge group, where $\chi$ is a new scalar singlet, $\Phi_2$ is an additional scalar doublet, and $N_{R_i}$ are right-handed neutrino fields. Importantly, the $B-L$ charges of these fields are distinct and crucial for the mechanism of neutrino mass generation.
\begin{table}[h]
\centering
\caption{BSM Field Content and Quantum Numbers}
\begin{tabular}{lcccc}
\toprule
Field & $SU(2)_L$ & $U(1)_Y$ & $U(1)_{B-L}$ & Spin \\
\midrule
$\chi$ & 1 & 0 & 2 & 0 \\
$\Phi_2$ & 2 & 1 & 0 & 0 \\
$N_{R_i}$ ($i=1,2,3$) & 1 & 0 & $-1$ & $\frac{1}{2}$ \\
\bottomrule
\end{tabular}
\label{tab:BSM_fields}
\end{table}

The complete Lagrangian of the model is the sum of the Standard Model Lagrangian $\mathcal{L}_{\text{SM}}$ and the new physics Lagrangian $\mathcal{L}_{\text{BSM}}$, which is responsible for the new interactions. The relevant part of the Lagrangian that couples the SM lepton doublets $L_i$ to the new right-handed neutrinos $N_{R_i}$ and the new scalar fields $\Phi_1$, $\Phi_2$, and $\chi$ is given by:
\begin{equation}
\mathcal{L}_y^{\text{BSM}} = -\left(
Y_{N_{ij}}^1 \bar{L}_i \tilde{\Phi}_1 N_{R_j} 
+ Y_{N_{ij}}^2 \bar{L}_i \tilde{\Phi}_2 N_{R_j} 
- \frac{1}{2} y_i \overline{N_{R_i}^C} \chi N_{R_i} 
+ \text{h.c.}
\right)
\end{equation}
where $\tilde{\Phi}_1 = i\sigma_2 \Phi_1^*$ is the conjugate scalar doublet. The first two terms describe Yukawa couplings between the SM lepton doublets and the new scalar doublets, $\Phi_1$ and $\Phi_2$, while the last term represents a Majorana mass term for the right-handed neutrinos, which is induced by the scalar singlet $\chi$. This structure is consistent with the seesaw mechanism, where the right-handed neutrinos obtain a large mass via the VEV of the scalar $\chi$, thereby providing a natural explanation for the small masses of the observed neutrinos.

The scalar potential of the model is given by:
\begin{eqnarray}
V(\Phi_1, \Phi_2, \chi) &= &\lambda_1 |\Phi_1|^4 + \lambda_2 |\Phi_2|^4 + \lambda_3 |\Phi_1|^2 |\Phi_2|^2 + \lambda_4 |\chi|^4 + \lambda_5 |\Phi_1^\dagger \Phi_2|^2 \nonumber\\
&+& \lambda_6 \left[ (\Phi_1^\dagger \Phi_2)^2 + \text{h.c.} \right] + \lambda_7 |\chi|^2 |\Phi_1|^2 + \lambda_8 |\chi|^2 |\Phi_2|^2
\end{eqnarray}
The coupling constants $\lambda_1$ through $\lambda_8$ control the strength of the interactions. The model respects several key symmetries:
\begin{itemize}

\item Classical Conformal Invariance: The model is constructed such that there are no explicit mass terms for the scalar fields. This symmetry is broken only when the fields acquire their vacuum expectation values (VEVs).

\item $\mathbb{Z}_2$ Symmetry: A discrete $\mathbb{Z}_2$ symmetry is imposed, under which the scalar doublet $\Phi_2$ and the right-handed neutrinos $N_R$ change sign, while all other fields remain invariant. This symmetry ensures that $\Phi_2$ remains inert, meaning it does not acquire a VEV, thus preventing spontaneous violation of the $B-L$ symmetry.

\item $U(1)_{B-L}$ Gauge Symmetry: The model maintains the new $U(1)_{B-L}$ gauge symmetry, which is essential for the mechanism of neutrino mass generation and the potential for dark matter candidates.
\end{itemize}

The vacuum structure of the model is crucial for understanding the breaking of both electroweak symmetry and the $B-L$ symmetry. The scalar fields acquire the following VEVs:
\begin{equation}
\langle \Phi_1 \rangle = \begin{pmatrix} 0 \\ v/\sqrt{2} \end{pmatrix}, \quad
\langle \chi \rangle = \frac{v_{B-L}}{\sqrt{2}}, \quad
\langle \Phi_2 \rangle = 0 \quad \text{(inert condition)}
\end{equation}
Here, $v$ is the usual electroweak scale, while $v_{B-L}$ is the VEV of the new scalar $\chi$, which sets the scale of the $B-L$ symmetry breaking. The condition $\langle \Phi_2 \rangle = 0$ enforces that $\Phi_2$ does not acquire a VEV, preserving the $\mathbb{Z}_2$ symmetry. The non-zero VEV of $\chi$ generates the Majorana mass terms for the right-handed neutrinos, thereby giving rise to neutrino masses through the seesaw mechanism.

%%%%%%%%%%%%%%%%%%%%%%%%%%%%%%%%%%%%%%%%%%%%
\section{First Order Phase Transition}
\
We consider two first order phase transitions (FOPT) in our model: (1) the $\chi$-driven FOPT I that can produce both stochastic gravitational wave backgrounds (SGWB) and primordial black holes (PBH), and (2) the $\Phi_{1,2}$-driven FOPT II that generates SGWB.
The complete one-loop corrected potential for the singlet scalar driven FOPT associated with U(1)$_{B-L}$ symmetry breaking is given by:
\begin{equation}
    V_{\text{tot}} = V_{\text{tree}} + V_{\text{CW}} + V_{\text{th}},
\end{equation}
where $V_{\text{tree}}$ is the tree-level potential. The Coleman-Weinberg potential $V_{\text{CW}}$ incorporates quantum corrections:
\begin{equation}
    V_{\text{CW}}(\phi) = (-1)^{2s_i} g_i \frac{m_i^4(\phi)}{64\pi^2} \left[ \log\left(\frac{m_i^2(\phi)}{\Lambda^2}\right) - c_i \right],
\end{equation}
with $s_i$, $g_i$, $m_i(\phi)$, $\Lambda$, and $c_i$ representing spin, degrees of freedom, field-dependent masses, renormalization scale, and scheme-dependent constants respectively.

The thermal potential $V_{\text{th}}$ includes finite temperature effects:
\begin{equation}
    V_{\text{th}}(\phi,T) = \frac{T^4}{2\pi^2} \left[ \sum_{\text{bosons}} J_B \left(\frac{m_i(\phi)}{T}\right) + \sum_{\text{fermions}} J_F \left(\frac{m_i(\phi)}{T}\right) \right] + V_D(\phi,T),
\end{equation}
where $J_{B/F}(y) = \int_0^\infty dx\, x^2 \ln(1 \mp e^{-\sqrt{x^2 + y^2}})$ are thermal loop functions (upper sign for fermions). The Debye correction $V_D$ accounts for resummation effects:
\begin{equation}
    V_D(\phi,T) = -\sum_i \frac{g_i T}{12\pi} \left[m_i^3(\phi,T) - m_i^3(\phi)\right],
\end{equation}
with thermal masses $m_i^2(\phi,T) = m_i^2(\phi) + \Pi_i(T)$.

The phase transition occurs when the effective potential develops a new minimum that becomes energetically favorable. The nucleation rate $\Gamma \approx T^4 e^{-S_3(T)/T}$ determines the transition dynamics, where $S_3(T)$ is the three-dimensional Euclidean action. The transition completes when $\int_{T_c}^T \frac{dT'}{T'} \frac{\Gamma(T')}{H^4(T')} \sim 1$, with $T_c$ being the critical temperature. The strength $\alpha$ and inverse duration $\beta$ of the phase transition, crucial for gravitational wave production, are given by:
\begin{equation}
    \alpha = \frac{\Delta V - T \Delta(dV/dT)}{4\rho_{\text{rad}}}, \quad \frac{\beta}{H_*} = T_* \left.\frac{d(S_3/T)}{dT}\right|_{T=T_*},
\end{equation}
where $\rho_{\text{rad}}$ is the radiation energy density and $T_*$ is the nucleation temperature.
%%%%%%%%%%%%%%%%%%%%%%%%%%%%%%%%%%%%%%%%%%%%%%

\section{GW Production from FOPT}

FOPTs generate stochastic GW backgrounds through bubble collisions, sound waves in the plasma, and magnetohydrodynamic turbulence. We consider two scenarios: strongly supercooled FOPTs where GWs come from bubble collisions and curvature perturbations, and weaker FOPTs dominated by sound waves.
For strongly supercooled FOPTs, the total GW spectrum combines collision and curvature contributions:
\begin{equation}
\Omega_{\mathrm{GW}}h^2 \approx 1.6\times 10^{-5}\left(\dfrac{g_*}{100}\right)\left(\Omega_{\mathrm{coll}} + \Omega_{\mathrm{curv}}\right)
\end{equation}
where the collision spectrum $\Omega_{\mathrm{coll}}$ scales with the inverse duration $\beta/H$ and peaks at wavenumber $k_p \approx 0.7 k_{\mathrm{max}}\beta/H$, while $\Omega_{\mathrm{curv}}$ depends on the curvature power spectrum $\mathcal{P}_\zeta$.

Sound wave contributions dominate for weaker transitions:
\begin{equation}
\Omega_{\mathrm{sw}}h^2 \approx 4.13\times 10^{-7}(R_*H_*)\left(\dfrac{\kappa_{\mathrm{sw}}\alpha}{1+\alpha}\right)^2 S_{\mathrm{sw}}(f)
\end{equation}
with peak frequency scaling as:
\begin{equation}
f_{\mathrm{sw}} \approx 2.6\times 10^{-5}\,\mathrm{Hz}\,(R_*H_*)^{-1}\left(\dfrac{T_{\mathrm{reh}}}{100\,\mathrm{GeV}}\right)
\end{equation}

The bubble wall velocity $v_w$ transitions between non-relativistic and ultra-relativistic regimes:
\begin{equation}
v_w = \begin{cases}
\sqrt{\Delta V_{\mathrm{tot}}/\alpha\rho_\mathrm{rad}} & \text{for slow walls}\\
1 & \text{for relativistic walls}
\end{cases}
\end{equation}
The GW spectrum depends critically on the transition strength $\alpha$, inverse duration $\beta/H$, bubble dynamics ($v_w$, $\kappa_{\mathrm{sw}}$), and thermal history ($T_{\mathrm{reh}}$, $g_*$). Strong FOPTs produce high-frequency GWs from collisions, while weaker transitions generate lower-frequency signals from sound waves, with remaining uncertainties in turbulence modeling.

%%%%%%%%%%%%%%%%%%%%%%%%%%%%%%%
\section{PBH Formation from FOPT}

Primordial black holes (PBHs) form during first-order phase transitions when delayed bubble nucleation creates energy density inhomogeneities between false vacuum regions and their true vacuum surroundings. Following \cite{Lewicki:2024ghw}, the characteristic PBH mass scale is set by the horizon mass at formation:
\begin{equation}
M_H = 10^{32}\left(\dfrac{100}{g_*}\right)^{1/2}\left(\dfrac{T_{\mathrm{reh}}}{\mathrm{GeV}}\right)^{-2}~\mathrm{g}
\end{equation}
The PBH abundance depends exponentially on the phase transition duration parameter $\beta/H$:
\begin{equation}
f_{\mathrm{PBH}} \approx 2.87\times 10^6 \exp\left(-0.07e^{0.754\beta/H}\right)\left(\dfrac{g_*}{g_{*s}}\right)\left(\dfrac{T_{\mathrm{reh}}}{\mathrm{GeV}}\right)
\end{equation}
The mass distribution follows a peaked profile with scaling parameters $\gamma = 0.36$, $c_1 \approx 1.5$, and $c_2 \approx 3.2$ for $\beta/H = 8$:
\begin{equation}
\psi(M) \propto (M/M_H)^{1+1/\gamma}\exp\left(-c_1(M/M_H)^{c_2}\right)
\end{equation}
The distribution peaks at $M_{\mathrm{PBH,peak}} \approx 0.928M_H$, with formation requiring critical overdensity $\delta_c = 0.5$ to overcome radiation pressure. The PBH properties are thus determined by the phase transition parameters $\beta/H$, $T_{\mathrm{reh}}$, and the effective degrees of freedom $g_*$.
%%%%%%%%%%%%%%%%%%%%%%%%%%%%%%%%%%%%
\section{Benchmark Scenarios and Phenomenological Outcomes}

We analyze two representative benchmark scenarios to demonstrate the observable consequences of our model. Table~\ref{bpin} specifies the input parameters for each case, where we have expressed the scalar sector through physical masses ($m_{H,A,H^{\pm}}$) and the combination $\lambda_{356}$.
\begin{table}[h]
\centering
\begin{tabular}{|c|c|c|c|c|c|c|c|c|c|}
\hline
BP & $g_{B-L}$ & $y_{1,2,3}$ & $v_{\chi}$ (GeV) & $\lambda_4$ & $\lambda_2$ & $m_H$ (GeV) & $m_A$ (GeV) & $m_{H^{\pm}}$ (GeV) & $\lambda_{356}$ \\ \hline
1 & 0.2004 & 0.1356 & $10^8$ & $10^{-10}$ & 2 & 500 & 1000 & 800 & 10 \\ \hline
2 & 0.2166 & 0.2146 & $10^7$ & $10^{-10}$ & 3 & 300 & 600 & 1000 & 5 \\ \hline
\end{tabular}
\caption{Input parameters for the benchmark scenarios.}
\label{bpin}
\end{table}

\begin{figure}[t]
\centering
\includegraphics[scale=0.6]{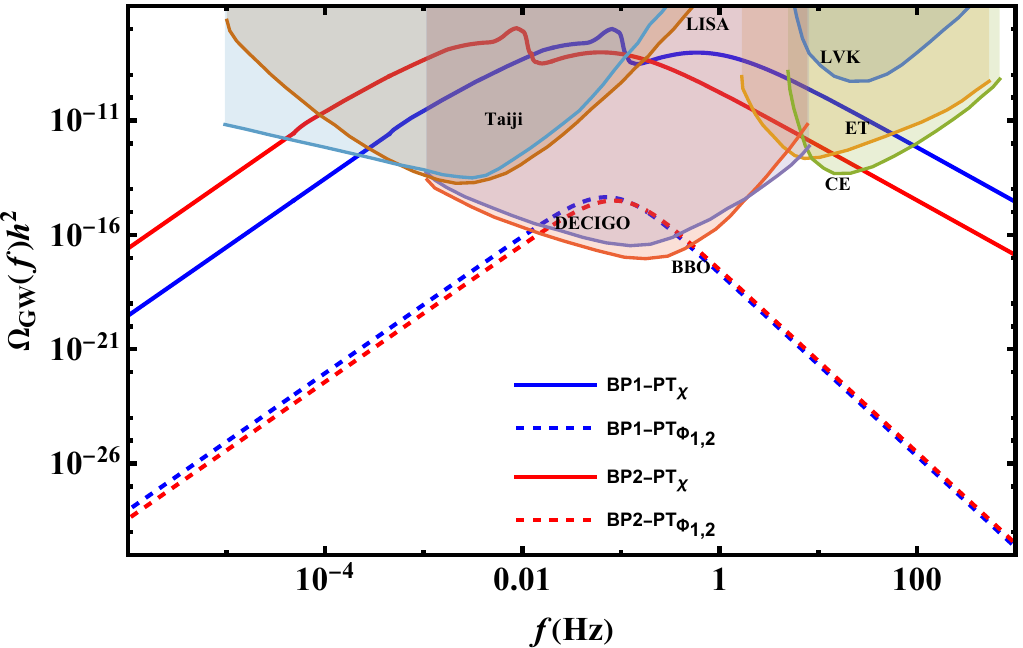}[h!]
\caption{Stochastic gravitational wave spectra for both benchmark scenarios, compared with detector sensitivity curves.}
\label{BPGW}
\end{figure}

The phenomenological consequences appear in Table~\ref{bpout}, showing that PBHs account for essentially all dark matter in both scenarios. Each case produces two distinct first-order phase transitions (FOPTs) at different temperatures, generating separate stochastic gravitational wave background (SGWB) spectra shown in Figure~\ref{BPGW}.
\begin{table}[h]
\centering
\begin{tabular}{|c|c|c|c|c|}
\hline
BP & $f_{\mathrm{PBH}}$ & $M_{\mathrm{PBH}}$ (g) & $\chi$-FOPT ($\alpha, \beta/H, T_{\mathrm{reh}}$ (GeV)) & $\Phi_{1,2}$-FOPT ($\alpha, \beta/H, T_n$ (GeV)) \\ \hline
1 & 0.9997 & $6.81\times10^{18}$ & (160, 8.04, $3.69\times10^6$) & (0.013, 3905, 161) \\ \hline
2 & 0.9999 & $5.9\times10^{20}$ & (39750, 7.93, $3.96\times10^5$) & (0.0088, 4850, 176.2) \\ \hline
\end{tabular}
\caption{Phenomenological outputs for the benchmark scenarios.}
\label{bpout}
\end{table}

The $\chi$-driven FOPT produces a bimodal SGWB spectrum detectable by LISA, Taiji, DECIGO, BBO, CE, and ET, with the left peak arising from curvature perturbations and the right peak from bubble wall collisions. The doublet-driven FOPT generates signals only within BBO and DECIGO sensitivity ranges. Notably, while particle dark matter contributes $\mathcal{O}(10^{-4})$ to the relic density, PBHs dominate the dark matter content in both scenarios.

%%%%%%%%%%%%%%%%%%%%%%%%%%%%%%%%%%
\section{Conclusion}  

We have studied a $U(1)_{B-L}$ extension of the inert doublet model, incorporating a scalar singlet $\chi$ and right-handed neutrinos. The model predicts two distinct first-order phase transitions (FOPTs): a high-temperature $\chi$-driven transition and a low-temperature transition from the Higgs doublets. The high-temperature FOPT generates primordial black holes (PBHs) that fully account for dark matter, with particle dark matter contributions being negligible ($\mathcal{O}(10^{-4})$). Both FOPTs produce stochastic gravitational wave backgrounds (SGWBs) detectable by future observatories (LISA, Taiji, DECIGO, etc.), though the $\chi$-driven signal dominates. The FOPT strength $\alpha$ decreases sharply with gauge coupling $g_{B-L}$ but depends weakly on Yukawa couplings $y_{1,2,3}$ and $v_\chi$, while PBH formation requires fine-tuned parameters, exhibiting extreme sensitivity to coupling variations. This work demonstrates how $U(1)$ extensions can revive constrained models by enabling PBH dark matter via strong, slow FOPTs, a mechanism applicable to other phenomenologically rich but observationally challenged scenarios.


\begin{thebibliography}{99}

\bibitem{Zeldovich:1967lct}
Y.~B.~Zel'dovich and I.~D.~Novikov,
%``The Hypothesis of Cores Retarded during Expansion and the Hot Cosmological Model,''
Sov. Astron. \textbf{10} (1967), 602

\bibitem{Carr:1974nx}
B.~J.~Carr and S.~W.~Hawking,
%``Black holes in the early Universe,''
Mon. Not. Roy. Astron. Soc. \textbf{168} (1974), 399-415
doi:10.1093/mnras/168.2.399

\bibitem{Niikura:2017zjd}
H.~Niikura, M.~Takada, N.~Yasuda, R.~H.~Lupton, T.~Sumi, S.~More, T.~Kurita, S.~Sugiyama, A.~More and M.~Oguri, \textit{et al.}
%``Microlensing constraints on primordial black holes with Subaru/HSC Andromeda observations,''
Nature Astron. \textbf{3} (2019) no.6, 524-534
doi:10.1038/s41550-019-0723-1
[arXiv:1701.02151 [astro-ph.CO]].

\bibitem{Griest:2013esa}
K.~Griest, A.~M.~Cieplak and M.~J.~Lehner,
%``New Limits on Primordial Black Hole Dark Matter from an Analysis of Kepler Source Microlensing Data,''
Phys. Rev. Lett. \textbf{111} (2013) no.18, 181302
doi:10.1103/PhysRevLett.111.181302

\bibitem{Niikura:2019kqi}
H.~Niikura, M.~Takada, S.~Yokoyama, T.~Sumi and S.~Masaki,
%``Constraints on Earth-mass primordial black holes from OGLE 5-year microlensing events,''
Phys. Rev. D \textbf{99} (2019) no.8, 083503
doi:10.1103/PhysRevD.99.083503
[arXiv:1901.07120 [astro-ph.CO]].

\bibitem{Macho:2000nvd}
R.~A.~Allsman \textit{et al.} [Macho],
%``MACHO project limits on black hole dark matter in the 1-30 solar mass range,''
Astrophys. J. Lett. \textbf{550} (2001), L169
doi:10.1086/319636
[arXiv:astro-ph/0011506 [astro-ph]].

\bibitem{EROS-2:2006ryy}
P.~Tisserand \textit{et al.} [EROS-2 Collaboration],
``Limits on the Macho Content of the Galactic Halo from the EROS-2 Survey of the Magellanic Clouds,''
\textit{Astron. Astrophys.} \textbf{469}, 387?404 (2007)
doi:10.1051/0004-6361:20066017
[arXiv:astro-ph/0607207].

\bibitem{Zumalacarregui:2017qqd}
M.~Zumalac\'arregui and U.~Seljak,
``Limits on stellar-mass compact objects as dark matter from gravitational lensing of type Ia supernovae,''
\textit{Phys. Rev. Lett.} \textbf{121}, no.14, 141101 (2018)
doi:10.1103/PhysRevLett.121.141101
[arXiv:1712.02240 [astro-ph.CO]].

\bibitem{Poulin:2017bwe}
V.~Poulin, P.~D.~Serpico, J.~Lesgourgues, and T.~Tram,
``Cosmological imprints of primordial black holes: A review,''
\textit{Phys. Rept.} \textbf{709}, 1?52 (2017)
doi:10.1016/j.physrep.2017.09.001
[arXiv:1707.04206 [astro-ph.CO]].

\bibitem{LIGOScientific:2016aoc}
B.~P.~Abbott \textit{et al.} [LIGO Scientific Collaboration and Virgo Collaboration],
``Observation of Gravitational Waves from a Binary Black Hole Merger,''
\textit{Phys. Rev. Lett.} \textbf{116}, no. 6, 061102 (2016)
doi:10.1103/PhysRevLett.116.061102
[arXiv:1602.03837 [gr-qc]].

\bibitem{Banerjee:2024fam}
I.~K.~Banerjee, U.~K.~Dey and S.~Khalil,
%``Primordial Black Holes and Gravitational Waves in the U(1)$_{B?L}$ extended inert doublet model: a first-order phase transition perspective,''
JHEP \textbf{12} (2024), 009
doi:10.1007/JHEP12(2024)009
[arXiv:2406.12518 [hep-ph]].

\bibitem{Lewicki:2024ghw}
M.~Lewicki, P.~Toczek and V.~Vaskonen,
%``Black holes and gravitational waves from slow phase transitions,''
Phys.Rev.Lett. 133 (2024) 22, 221003
doi:10.1103/PhysRevLett.133.221003
[arXiv:2402.04158 [astro-ph.CO]]



\end{thebibliography}
\end{document}